\documentclass[preprint,floatfix,onecolumn,superscriptaddress]{revtex4-1}
\usepackage[english]{babel}
\usepackage{amsmath}
\usepackage{amssymb}
\usepackage{tensor}
\usepackage{graphicx}
\usepackage{subfigure}

\newcommand{\be}{\begin{eqnarray}}
\newcommand{\ee}{\end{eqnarray}}

\newcommand{\dc}{c^{\dagger}}

\begin{document}
	
\title{Mobility edges in one dimensional models with large quasi-periodic disorders}

\author{Qiyun Tang}
\affiliation{College of Physics, Sichuan University, Chengdu, Sichuan 610064, China}

\author{Yan He}
\affiliation{College of Physics, Sichuan University, Chengdu, Sichuan 610064, China}
\email{heyan$_$ctp@scu.edu.cn}

\begin{abstract}
We study the one-dimensional tight-binding model with quasi-periodic disorders, where the quasi-period is tuned to be very large. It is found that this type of model with large quasi-periodic disorders can also support the mobility edges, which is very similar to the models with slowly varying quasi-periodic disorders. The energy matching method is employed to determine the locations of mobility edges in both types of models. These results of mobility edges are verified by numerical calculations in various examples. We also provide a qualitative arguments to support the fact that large quasi-periodic disorders will lead to the existence of mobility edges.

\end{abstract}

\maketitle

\section{Introduction}

The phenomena of localization \cite{Anderson} has been extensively studied in the condensed matter physics for many years. For three-dimensional (3D) systems, there exists a threshold of disorder strength, above which the Anderson localization will take place. For generic band structures, a closely related notion is the mobility edge \cite{Mott} which is the critical energy separating the localized and extended energy eigenstates. Since the direct observation of mobility edges in 3D system is quite challenging, the attentions have been turned to low-dimensional systems. However, the number of dimension is a very important factor for Anderson localization. It is argued by the scaling theory \cite{scaling1,scaling2} that for one-dimensional (1D) systems even infinitesimal small disorders will make all eigenstates complete localized. Thus, one cannot find any the mobility edge in 1D band stucture.

About forty years ago, the low-dimensional quasi-periodic systems with correlated disorders have captured a lot of attentions ever since.  The most famous example is the so called Aubry-Andre (AA) model \cite{Aubry,Harper}. In essence, the AA model is just a 1D tight-binding model with incommensurate or quasi-periodic on-site potentials. By Fourier transformation, one can show that the hopping terms and potential terms of AA model can transfer into each other. By equating the hopping terms with the potential terms, one can see that there exists a self-dual symmetry in AA model. It is easy to see that this self-dual point will separate between the localized and extended states. Unfortunately, one still cannot find any mobility edges in the standard AA model.

In order to make mobility edges possible in the AA type models, Das Sarma and co-workers \cite{Xie1,Xie2,Biddle} introduce the slowly varying quasi-periodic potentials into the AA model. This new class of 1D model was shown to support the mobility edges and has been extensively studied ever since. The success of slowly varying quasi-periodic model stimulated a lot of more works on the mobility edges in 1D quasi-periodic systems \cite{1D1,1D2,1D3,1D4}. It is also found that the mobility edge can appear in many different types of modified AA model such as in \cite{hopping1,hopping2,hopping3,hopping4,hopping5,hopping6}. During the same time, the mobility edges of many different models have been studied \cite{M1,M2,M3,M4}. A variety of novel physical phenomena induced by the mobility edge and methods to determine the mobility edges are also explored extensively \cite{P1,P2,P3,P4,P5,P6}. In addition, the research of mobility edge has been further extended to the domain of non-Hermitian systems\cite{NH1,NH2,NH3,NH4,NH5,NH6,NH7}.

In this paper, we present another version of modified AA model which can also accommodate mobility edges. In this modified AA model, we simply replace the irrational frequency $\alpha$ by another much smaller irrational frequency such as $\alpha/M$ where $M\sim10^2$. Due to this much smaller irrational frequency, our model have a long quasi-period, thus we can call this model as ``AA model with large quasi-periodic disorders''. We will present a detailed study of mobility edges of this large quasi-periodic model with disorders in on-site potential or in hopping constant or in both. We will see that in each case, their mobility edges are very similar to the AA model with slow varying quasi-periodic disorders.

We can determine the location of these mobility edges by the so-called energy matching method which we have proposed in our previous works \cite{Tang-21}. In essence, we can approximate the above quasi-disordered model by an ensemble of different periodic models. Then the region of extended states can be approximately obtained by the overlaps of energy bands of all these periodic models. We will demonstrate this method works quite efficiently in determining the mobility edges for different versions of AA model with large quasi-periodic disorders. In the end of this paper, we will present a qualitative arguments in support the existence of mobility edges in the models with large quasi-periodic disorders

The rest of this paper is organized as follows. In section \ref{sec-AA}, we introduce the two types of quasi-periodic disordered models which will be studied closely. Then we present a brief review of the energy matching method in section \ref{sec-energy}. In section \ref{sec-num}, we will apply this method to determine the mobility edges of various disordered models and also use detailed numerical calculations to verify them. At last in section \ref{sec-explain}, we provide a qualitatively explanation of the existence of mobility edges in large quasi-periodic disordered models

\section{The large quasi-period disordered model and slowly varying disordered model}
\label{sec-AA}

In this paper, we will manly focus on the modified AA models with very small irrational frequencies or large quasi-periodic disorders.  The Hamiltonian of this type of models can be summarize as
\be
H=-\sum_{i=1}^{N-1}(t+w_i)(\dc_i c_{i+1}+\dc_{i+1} c_i)+\sum_{i=1}^N\mu_i \dc_i c_i
\label{AAH}
\ee
here $c_i(\dc_i)$ is the fermionic annihilation (creation) operator, $N$ is the total number of lattice sites. For convenience, we set $t=1$ as the energy unit. Disorders with large quasi-periods is introduced into the above model by the following assumptions
\be
w_i=w\cos\Big(2\pi \frac{\alpha}{M} i\Big),\quad
\mu_i=\mu\cos\Big(2\pi \frac{\beta}{M} i+\phi\Big)
\label{w}
\ee
Here $w$ and $\mu$ is the amplitude of the disorder and $i$ labels the lattice site. $\alpha$ and $\beta$ are certain order 1 irrational numbers, which determine the quasi-periodic behavior. In particular, if one set $\alpha=0$, then the model only contains disorders in potential terms. Similarly, one can also put the disorders in hopping term or in both hopping and potential terms. Here $\phi$ is some fixed phase angle, which can tune the relative disorders between the potential terms and hopping terms. Most importantly, we also introduce a large integer $M$ in the denominators, which is used to control the order of magnitude of irrational frequencies. It is easy to see that the length of quasi-period of $w_i$ and $\mu_i$ is given by
\be
T_{\alpha}=\frac{M}{\alpha}\qquad T_{\beta}=\frac{M}{\beta}
\label{T}
\ee

Throughout the whole paper, we will often compare the models of large quasi-periods disorders with the models of slowly varying quasi-periodic disorders. As we discussed in the introduction, the model with slowly varying quasi-period is well known to support the mobility edges. Its Hamiltonian can be expressed as
\be
H=-\sum_{i=1}^{N-1}(t+w'_i)(\dc_i c_{i+1}+\dc_{i+1} c_i)+\sum_{i=1}^N\mu'_i \dc_i c_i
\label{AAH'}
\ee
where
\be
w'_i=w'\cos(2\pi\alpha i^v),\quad
\mu'_i=\mu'\cos(2\pi\beta i^v+\phi)
\label{w'}
\ee
One usually assume the exponent $v$ to satisfy $0<v<1$. In this case, one can see both $w'_i$ and $\mu'_i$ approaches to some constant value as $i$ become very large. This behavior justifies the name of slowly varying quasi-period.

Due to the characteristic that $w'_i$ and $\mu'_i$ approach constants in the large $i$ limit, one can give a heuristic argument to determine the mobility edge of the slowly varying quasi-periodic model in certain range. In this method, the asymptotic constancy of $w'_i$ and $\mu'_i$ is considered to be an important condition for the generation of mobility edges. The large quasi-periodic model doesn't have this property that disorders gradually approaching constants, but it still produces mobility edges similar to the slowly varying quasi-periodic models. This shows that the asymptotic constancy of $w'_i$ and $\mu'_i$ is not a necessary condition for the existence of mobility edges.

There is a simple physical picture which can roughly explain why the mobility edges can exist in the slowly varying disordered models. For small $i$, the model of Eq.(\ref{AAH'}) will favor the localized states due to the disorders. On the other hand, for large $i$, the model of Eq.(\ref{AAH'}) becomes uniform and favors extended state. It is the competing of these two effects give rise to the mobility edges in the slowly varying disordered models. In the following discussions, we will see that the large quasi-periodic disordered models have very similar features as the slowly varying models.

In the rest of this paper, we will set the irrational number $\alpha$ or $\beta$ in the above two types of models to be $\frac{\sqrt{5}-1}2$. The total number of lattice sites to set to be $N=10000$.

\section{Introducing the energy matching method}
\label{sec-energy}

The ``energy matching method'' is an effective method to determine the mobility edge. The basic idea of this method is to approximate the quasi-periodic disordered models by an ensemble of periodic models. For convenience, we label these periodic models by $a=1,\cdots,N$.  Then their Hamiltonian can be written as
\be
&&H_a=-\sum_{i=1}^{N-1}(t+w_a)(\dc_i c_{i+1}+\dc_{i+1} c_i)+\sum_{i=1}^{N}\mu_a \dc_i c_i.\\
&&w_a=w\cos(2\pi \frac{\alpha}{M} a),\quad
\mu_a=\mu\cos(2\pi \frac{\beta}{M} a+\phi)\nonumber
\ee
Here periodic boundary condition such that $c_{N+1}=c_1$ is imposed and $w_a$ and $\mu_a$ do not depend lattice site. For each $H_a$, we can diagonalize the Hamiltonian in the momentum space to find the extended eigenstates with the following energy bands.
\be
E=\mu_a-2(t+w_a)\cos k
\label{E}
\ee
Although we can not diagonalize the quasi periodic disordered model analytically, one can expect intuitively that the eigenstates of a given energy $E$ is extended, if this energy $E$ is located inside in the energy band of all the periodic model $H_a$. Therefore, the energy region of extended state of the model of Eq.(\ref{AAH}) will be the overlap of all energy bands of Eq.(\ref{E})
\be
E\in \bigcap_a\Big(\mu_a-2(t+w_a),\,\mu_a+2(t+w_a)\Big)
\label{E1}
\ee

The energy matching method can also be applied to the model with slowly varying quasi periodic disorders. In this case, on only need to change $w_a$ and $\mu_a$ to the following
\be
w_a=w\cos(2\pi\alpha a^v),\quad
\mu_a=\mu\cos(2\pi\beta a^v+\phi)
\ee
In previous paper, we have verified that the energy matching method can give the correct location of mobility edges in this case.

In the next section, we will apply this method to calculate the mobility edges of the models with large quasi-period  and the models with slowly varying disorders in several different examples. We will verify that the numerical results of mobility edges are consistent with the result of energy matching method. In the meantime, the above two types of models will give rise the same mobility edges.

\section{Numerical verification of Mobility edges}
\label{sec-num}

In this section, we choose the inverse participation ratio (IPR) and Lyapunov exponent \cite{Thouless,Kohmoto,Schreiber}, which are common physical quantities in the disordered system, as the indicators to distinguish between the extended state and the local state. These indicators can help us to numerically verify the accuracy of the mobility edges which are obtained by the energy matching method in the previous section.

The inverse participation ratio (IPR) of the $n$-th normalized eigen-wave function $\psi_n$ is defined as
\be
\mbox{IPR}_n=\sum_{j=1}^N\Big|a^n_j\Big|^4,\quad
\psi_n=(a^n_1,\cdots,a^n_N)
\ee
For the extend states, one expects that their amplitudes are roughly uniform. Thus each component of $\psi_n$ is roughly the same order of magnitude $|a^n_j|\sim 1/N$ for all $j$. Then it is easy to see that the IPR of the extended state should be the order of $\sum_{j=1}^N\frac{1}{N^2}=\frac1N$, which approachs 0 as the system size $N\to\infty$.  One the other hand, the non-zero amplitude of localized states will mostly confined to only a few components, therefore in this case, we expect that the IPR will be order 1.

In the same time, one also calculated the Lyapunov exponent of the system, which is defined as
\be
\gamma(E_u)=\frac{1}{N-1}\sum_{u \neq v}^{N}\ln\lvert E_u-E_v \rvert -\frac{1}{N-1}\sum_{i}^{N-1}\ln\lvert t+w_{i}\rvert
\ee
It is well-known that the Lyapunov exponent is the inverse of the localization length. Therefore, one expects the non-zero Lyapunov exponent $\gamma \neq 0$ for the localized states. One the other hand, the Lyapunov exponent will close to zero for extended states.

\begin{figure}
	\centering
	\subfigure[]{
		\begin{minipage}[t]{0.4\linewidth}
			\centering
			\includegraphics[width=\textwidth]{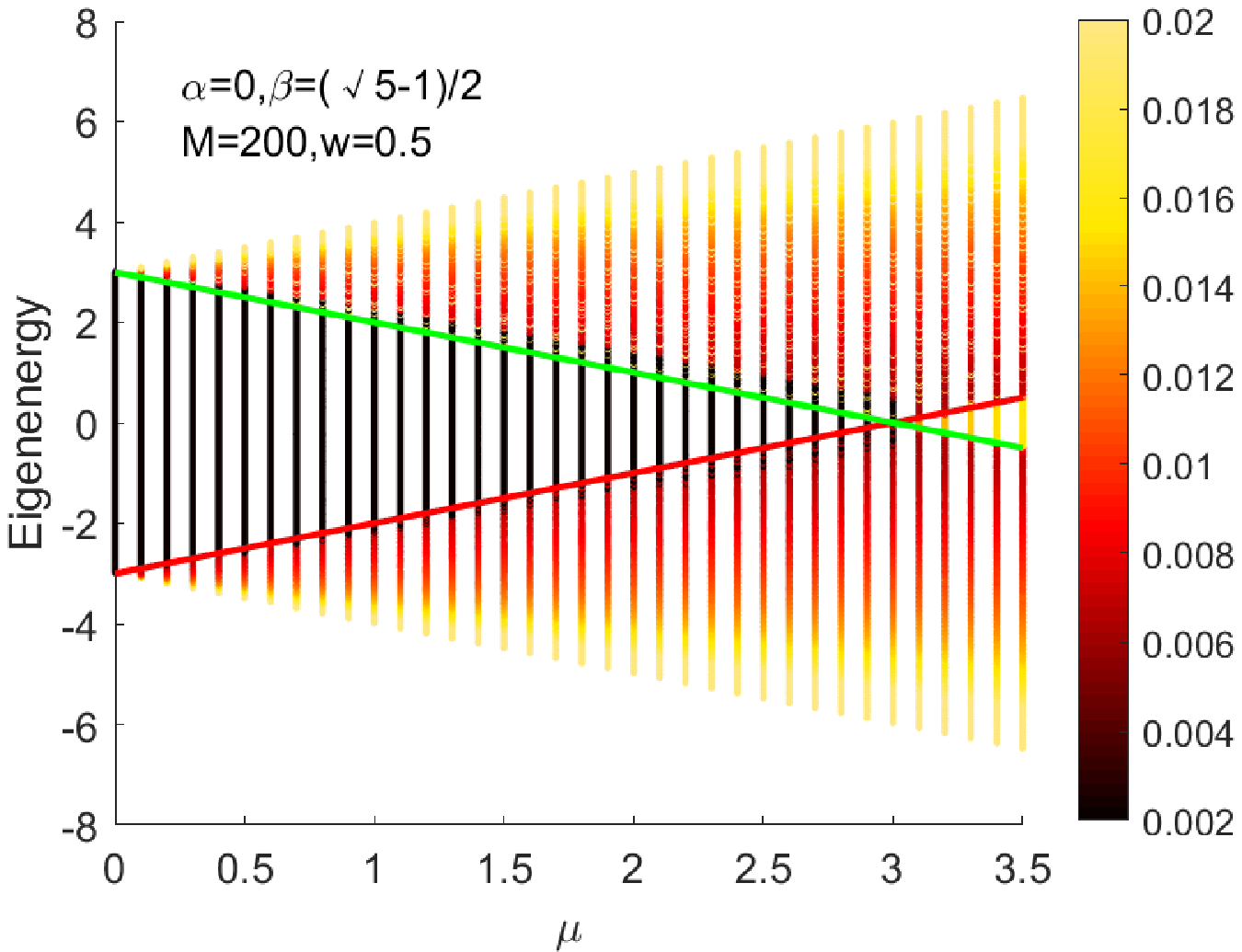}
		\end{minipage}
	}
	\subfigure[]{
		\begin{minipage}[t]{0.4\linewidth}
			\centering
			\includegraphics[width=\textwidth]{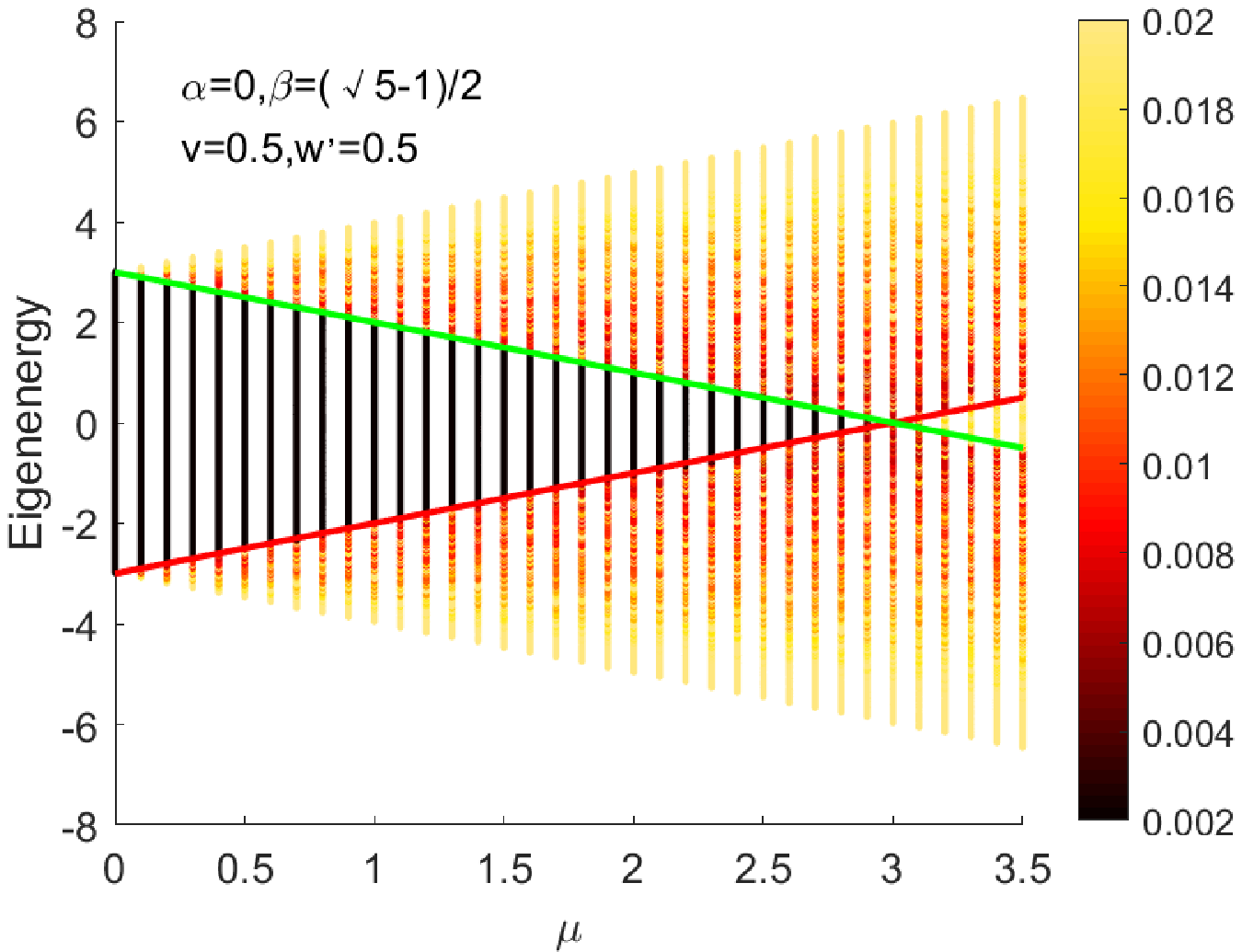}
		\end{minipage}
	}
	\quad
	\subfigure[]{
		\begin{minipage}[t]{0.4\linewidth}
			\centering
			\includegraphics[width=\textwidth]{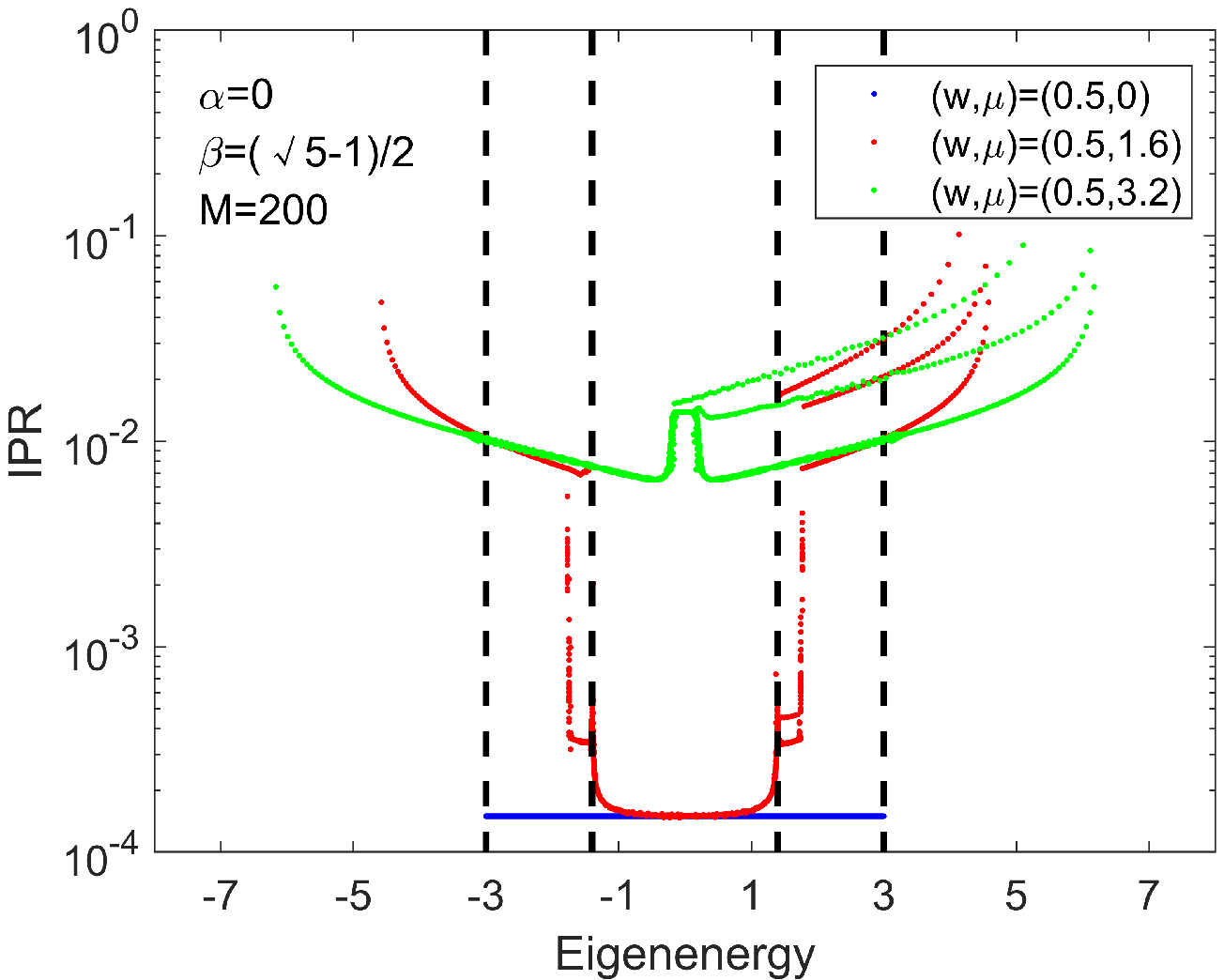}
		\end{minipage}
	}
	\subfigure[]{
		\begin{minipage}[t]{0.4\linewidth}
			\centering
			\includegraphics[width=\textwidth]{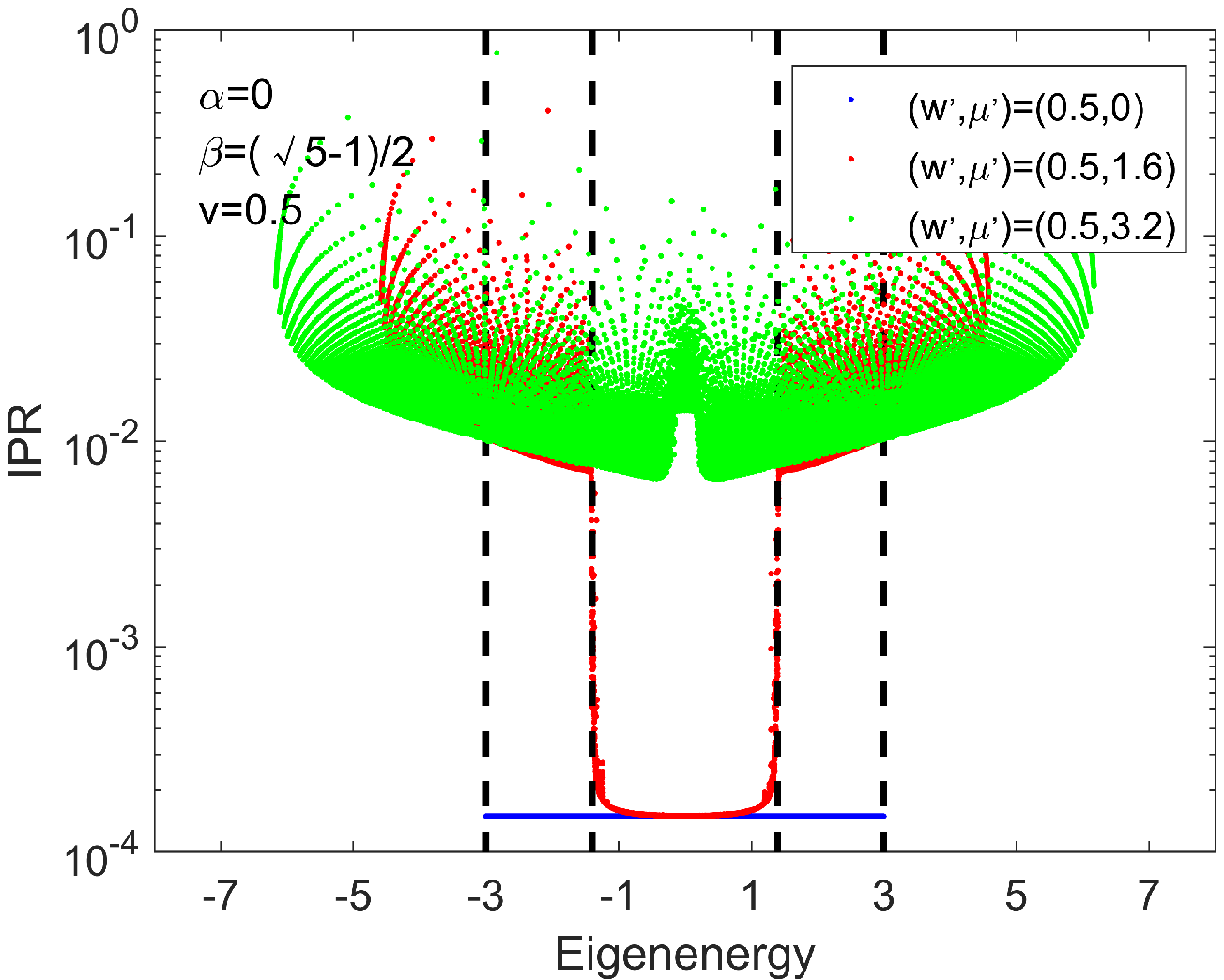}
		\end{minipage}
	}
	\subfigure[]{
		\begin{minipage}[t]{0.4\linewidth}
			\centering
			\includegraphics[width=\textwidth]{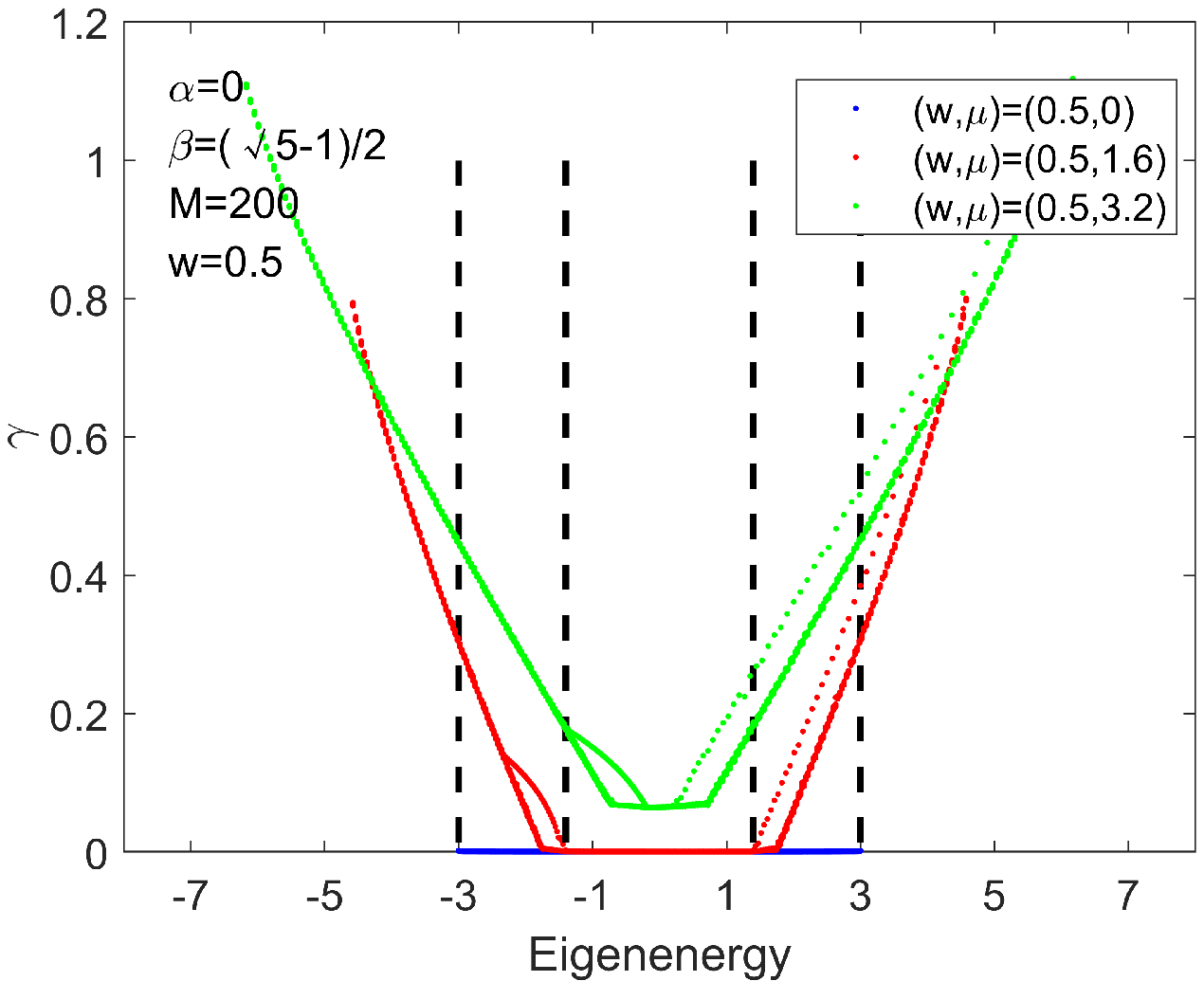}
		\end{minipage}
	}
	\subfigure[]{
		\begin{minipage}[t]{0.4\linewidth}
			\centering
			\includegraphics[width=\textwidth]{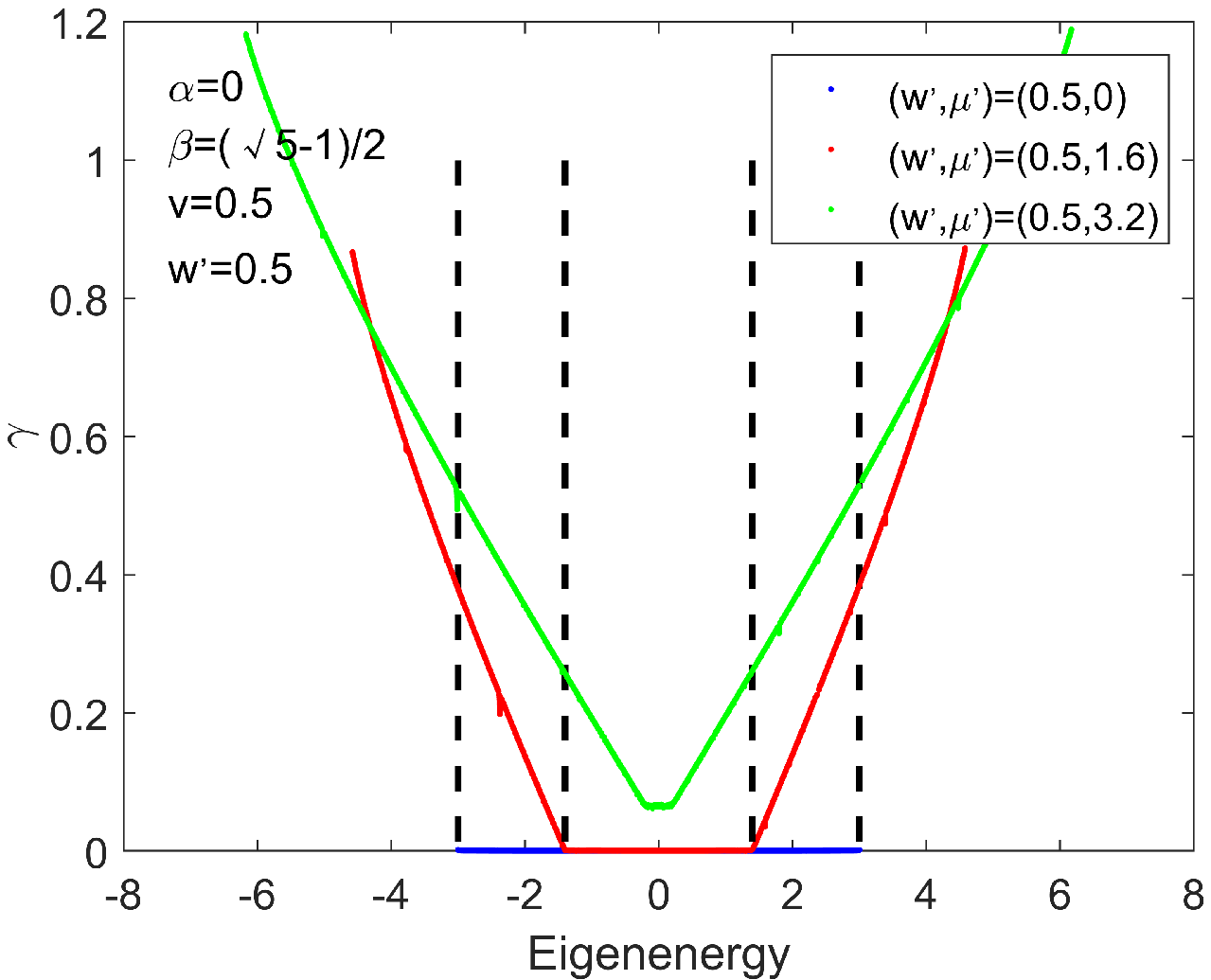}
		\end{minipage}
	}
	\centering
\caption{(a): Eigenenergy of the model of Eq.(\ref{AAH}) as a function of $\mu$ with $\alpha=0$, $\beta=(\sqrt{5}-1)/2$, $w=0.5$ and $M=200$.  (b): Eigenenergy of the model of Eq.(\ref{AAH'}) as a function of $\mu'$ with $w'=w$ and $v=0.5$, other parameters are the same as (a). The colors in panels (a) and (b) represent the IPR of each eigenstates. The red and green curves represent the mobility edges. (c) and (e): The IPR and Lyapunov exponent of Eq.(\ref{AAH}) as a function of eigenenergy for $(w,\mu)=(0.5,0),\,(0.5,1.6),\,(0.5,3.2)$, other parameters are the same as (a). (d) and (f): The IPR and Lyapunov exponent of Eq.(\ref{AAH'}) with $w'=w$ and $v=0.5$, other parameters are the same as (c).}
\label{AAH_3}
\end{figure}

\textbf{AA model with disordered potential}

As the first example, we consider the case with $\alpha=0$ and $\beta=(\sqrt{5}-1)/2$, which means the disorders are only in the potential. This is also the original version of AA model. In this case, $w_a$ is just a constant $w$. According to energy matching method, the lower bound of the region of extended states is determined by the maximum value of the bottom of each energy bands of $H_a$, which can be denoted as $E_{\text{bottom}}=\mu_a-2(t+w)$. Similarly, the upper bound of extended states is determined by the minimum value of the top of each energy band of $H_a$, which is $E_{\text{top}}=\mu_a+2(t+w)$.  It is easy to see that the resulting energy region of extended states is
\be
E\in \Big(\mu-2(t+w),\,-\mu+2(t+w)\Big)
\label{edge2}
\ee
As $\mu$ is tuned, the above upper and lower bound gives to lines of mobility edges separating extended states from localized states.

To verify the above mobility edges, we plot the eigenenergy of AA model with large quasi-period disordered potential of Eq.(\ref{AAH}) as a function of $\mu$ in Figure \ref{AAH_3} (a). The parameters used in this calculation is listed in the figure caption. The color of each points represents the IPR value of the eigenstates. One can see that the boundary of the dark region matches the mobility edges that we have decided above by the energy matching method. Similarly, we plot the eigenenergy of the slowly varying AA model Eq.(\ref{AAH'}) with disordered potential in Figure \ref{AAH_3} (b). The parameters of $\alpha$, $\beta$ in the slowly varying AA model are the same as those in the AA model with large quasi-period. We also assume $w=w'$ and the slowly varying exponent as $v=0.5$. It can be seen from the figure that the mobility edges of the two types of models are the same, and they are both consistent with the calculated results of Eq.(\ref{edge2}).

In order to make a more quantitatively observation, in Figure \ref{AAH_3} (c), (d), (e) and (f), we also plot the IPR values and Lyapunov exponent as a function of eigenenrgy for a few selected points as $\mu=0.9,\,1.6,\,2.5$. While figure \ref{AAH_3} (c), (d) shows the results of AA model with large quasi-period, and Figure \ref{AAH_3} (e), (f) shows the results of slowly-varying AA model. One can see that the IPR drops from the order of magnitude of $10^{-2}$ to $10^{-4}$ when the eigenenergy across certain critical values,  and the Lyapunov exponent will also change from non-zero to zero. This clearly signals the transition from localized states to extended states. These figures also support the conclusion that the mobility edges of the two types of models are both consistent with formula Eq.(\ref{edge2}).

\begin{figure}
	\centering
	\subfigure[]{
		\begin{minipage}[t]{0.45\linewidth}
			\centering
			\includegraphics[width=\textwidth]{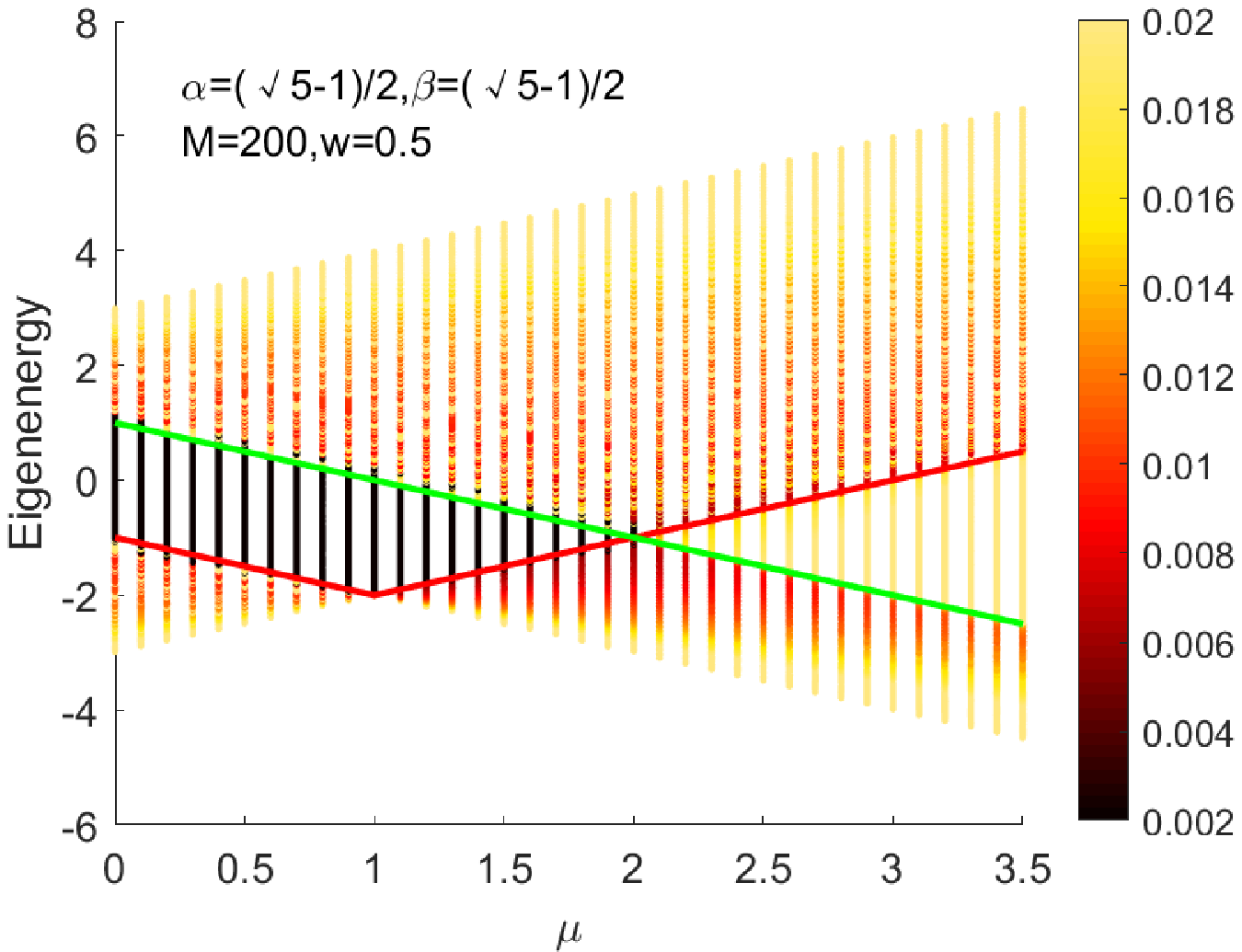}
		\end{minipage}
	}
	\subfigure[]{
		\begin{minipage}[t]{0.45\linewidth}
			\centering
			\includegraphics[width=\textwidth]{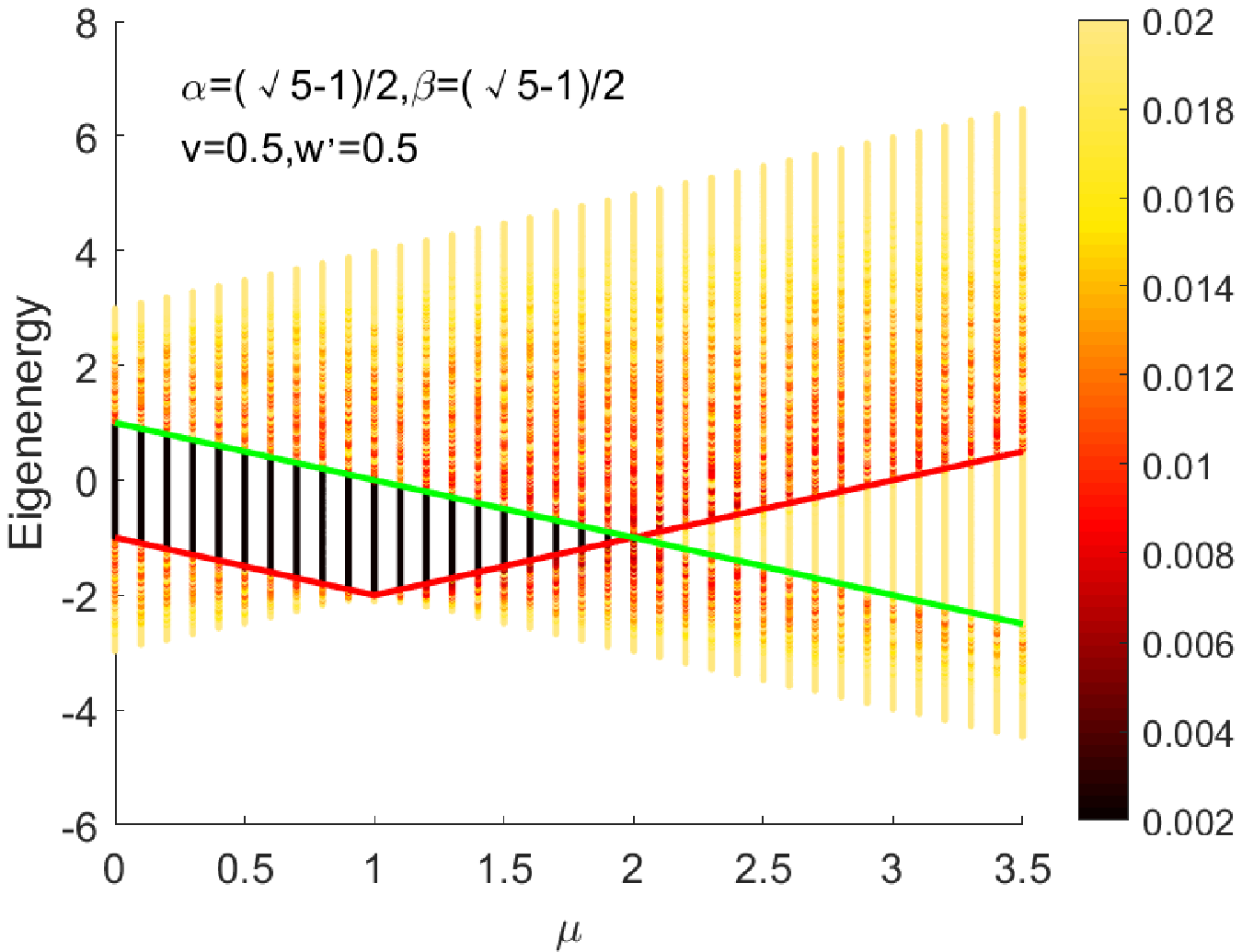}
		\end{minipage}
	}
	\quad
	\subfigure[]{
		\begin{minipage}[t]{0.45\linewidth}
			\centering
			\includegraphics[width=\textwidth]{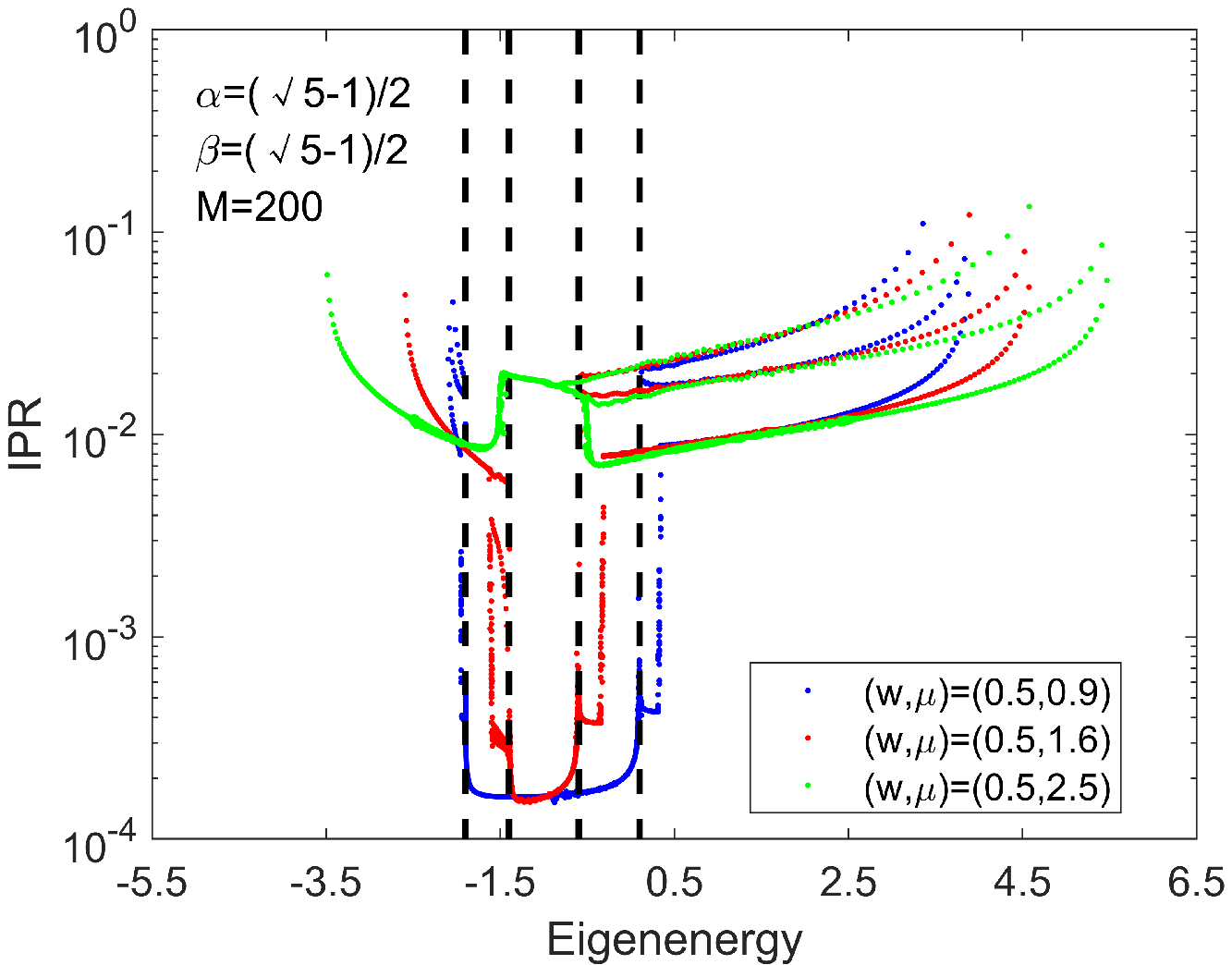}
		\end{minipage}
	}
	\subfigure[]{
		\begin{minipage}[t]{0.45\linewidth}
			\centering
			\includegraphics[width=\textwidth]{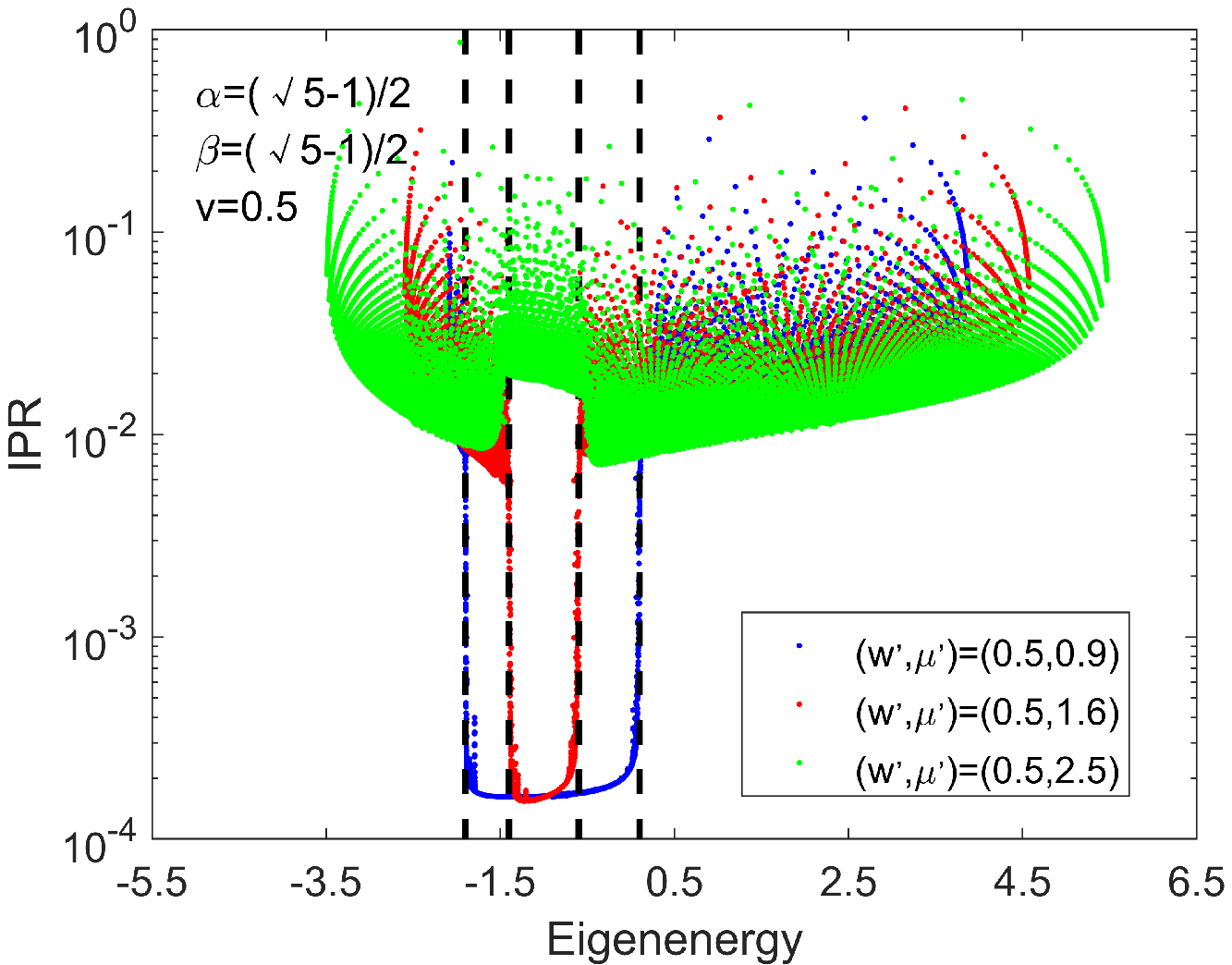}
		\end{minipage}
	}
\centering
\caption{(a): Eigenenergy of AA model Eq.(\ref{AAH}) as a function of $\mu$ with $\alpha=\beta=(\sqrt{5}-1)/2$, $w=0.5$ and $M=200$.  (b): Eigenenergy of slowly varying AA model Eq.(\ref{AAH'}) as a function of $\mu'$ with $w'=w$ and $v=0.5$, other parameters are the same as (a). The colors in panels (a) and (b) represent the IPR of each eigenstates. The red and green curves represent the mobility edges. (c): The IPR of AA model as a function of eigenenergy for $(w,\mu)=(0.5,0.9),\,(0.5,1.6),\,(0.5,2.5)$, other parameters are the same as (a). (d): The IPR of slowly varying AA model with $w'=w$ and $v=0.5$, other parameters are the same as (c).}
\label{AAH_2}
\end{figure}

\textbf{AA model with the same disorders in hopping and potential}

Now we turn to a more complicated model with $\beta=\alpha$ and $\phi=0$.  In this case, $\mu_i$ and $w_i$ have the same disorder. For this model, we can also use the energy matching method to determine the mobility edge. The lower bound of the energy region Eq.(\ref{E1}) of the extended state is determined by the extreme values $E=\mu-2(t+w)$ and $E=-\mu-2(t-w)$. Depending on the value of $\mu$, we will take the larger one as the lower bound. When $\mu>2w$, clearly, the lower bound of the energy region of the extended state is $E_{\min}=\mu-2(t+w)$. One the other hand, if $\mu<2w$, we have $E_{\min}=-\mu-2(t-w)$ as lower bound.

Similarly, we can also get extreme values $E=\mu+2(t+w)$ and $E=-\mu+2(t-w)$ for the upper bound of the energy region of the extended state. Obviously, regardless of the value of $\mu$ and $w$, the energy $E=-\mu+2(t-w)$ is smaller. So the upper bound of the energy region of the extended states is $E_{\max}=-\mu+2(t-w)$. In summary, when $\beta=\alpha$, we find the condition for extended states is given by
\be
&&E\in\Big(-\mu-2(t-w),\,-\mu+2(t-w)\Big),\quad \hbox{for}\, 0<\mu<2w\nonumber\\
&&E\in\Big(\mu-2(t+w),\,-\mu+2(t-w)\Big),\quad \hbox{for}\,  \mu>2w
\label{edge1}
\ee
To verify the above mobility edges, we show the eigenenergy of AA mdoel with large quasi-period in both hopping and potential in Figure \ref{AAH_2} (a). The parameters used in this calculation is listed in the figure caption. Again, the color of each points represents the IPR value of the eigenstates. Similarly, we also plot the eigenenergy of the slowly varying AA model with the same disorders in both hopping and potential in figure \ref{AAH_2} (b).  The values of the parameters of the slowly varying AA model are the same as those in the AA model with large quasi-period, but with the exponent $v=0.5$. It can be seen from the figure that the mobility edges of the two types of models are both consistent with the calculated results eq.(\ref{edge1}). In Figure \ref{AAH_2} (c), (d), the IPR values as a function of eigenenergy is shown for a few points $\mu=0.9,\,1.6,\,2.5$. Where panel (c) shows the results of AA model with large quasi-period, and panel (d) shows the results of slowly-varying AA model. It provides a more accurate verification of the mobility edges of Eq.(\ref{edge1}).

\begin{figure}
	\centering
	\subfigure[]{
		\begin{minipage}[t]{0.45\linewidth}
			\centering
			\includegraphics[width=\textwidth]{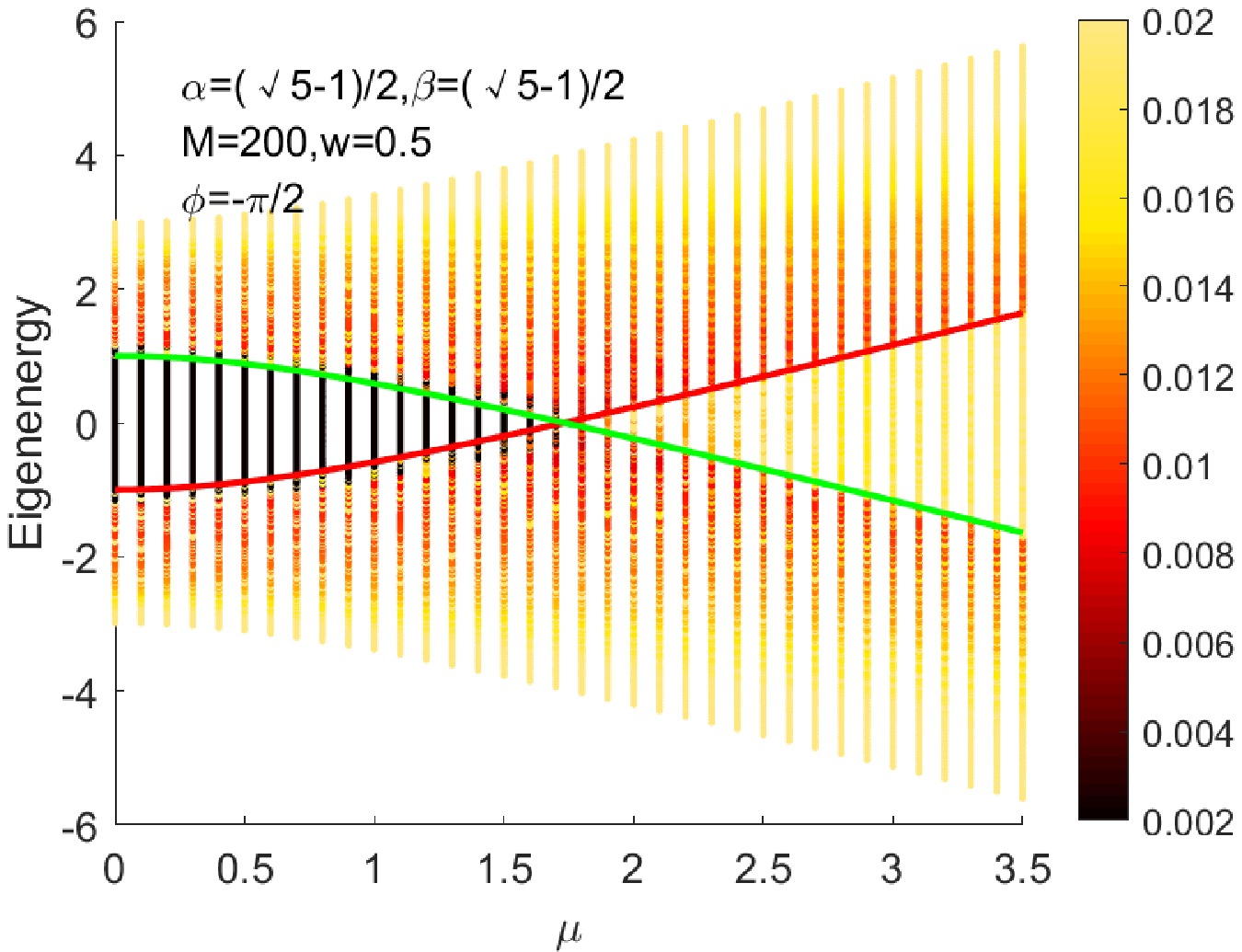}
		\end{minipage}
	}
	\subfigure[]{
		\begin{minipage}[t]{0.45\linewidth}
			\centering
			\includegraphics[width=\textwidth]{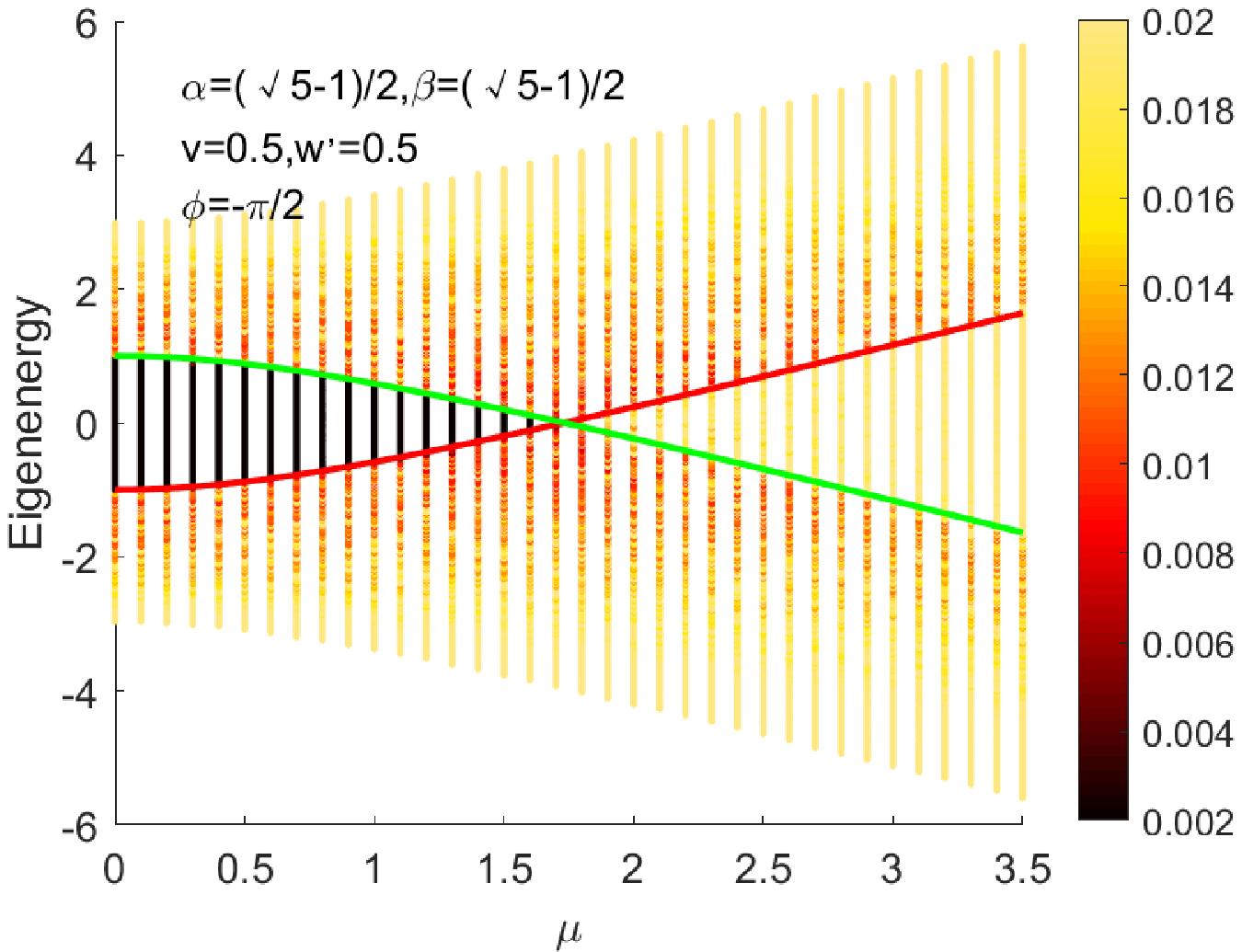}
		\end{minipage}
	}
	\quad
	\subfigure[]{
		\begin{minipage}[t]{0.45\linewidth}
			\centering
			\includegraphics[width=\textwidth]{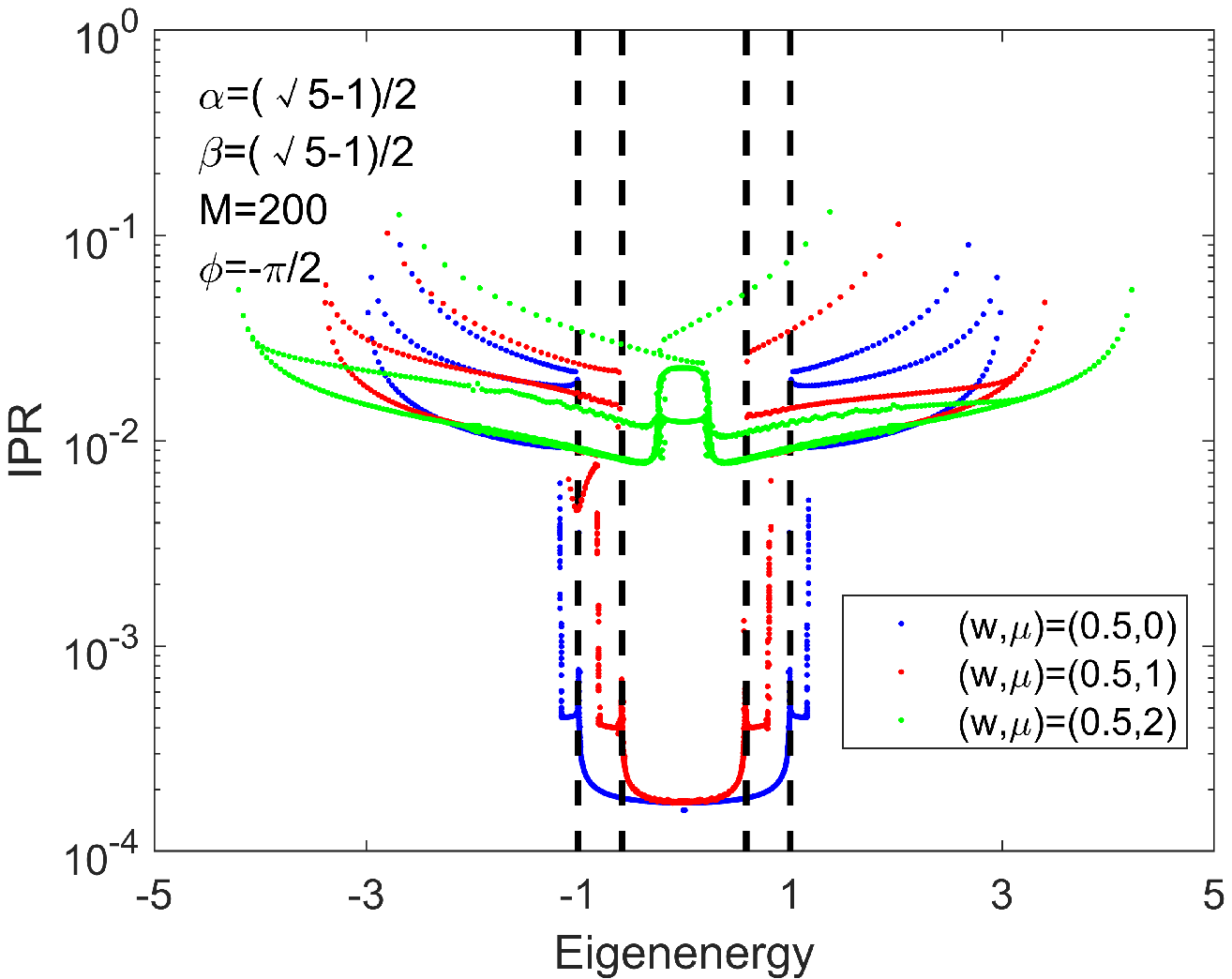}
		\end{minipage}
	}
	\subfigure[]{
		\begin{minipage}[t]{0.45\linewidth}
			\centering
			\includegraphics[width=\textwidth]{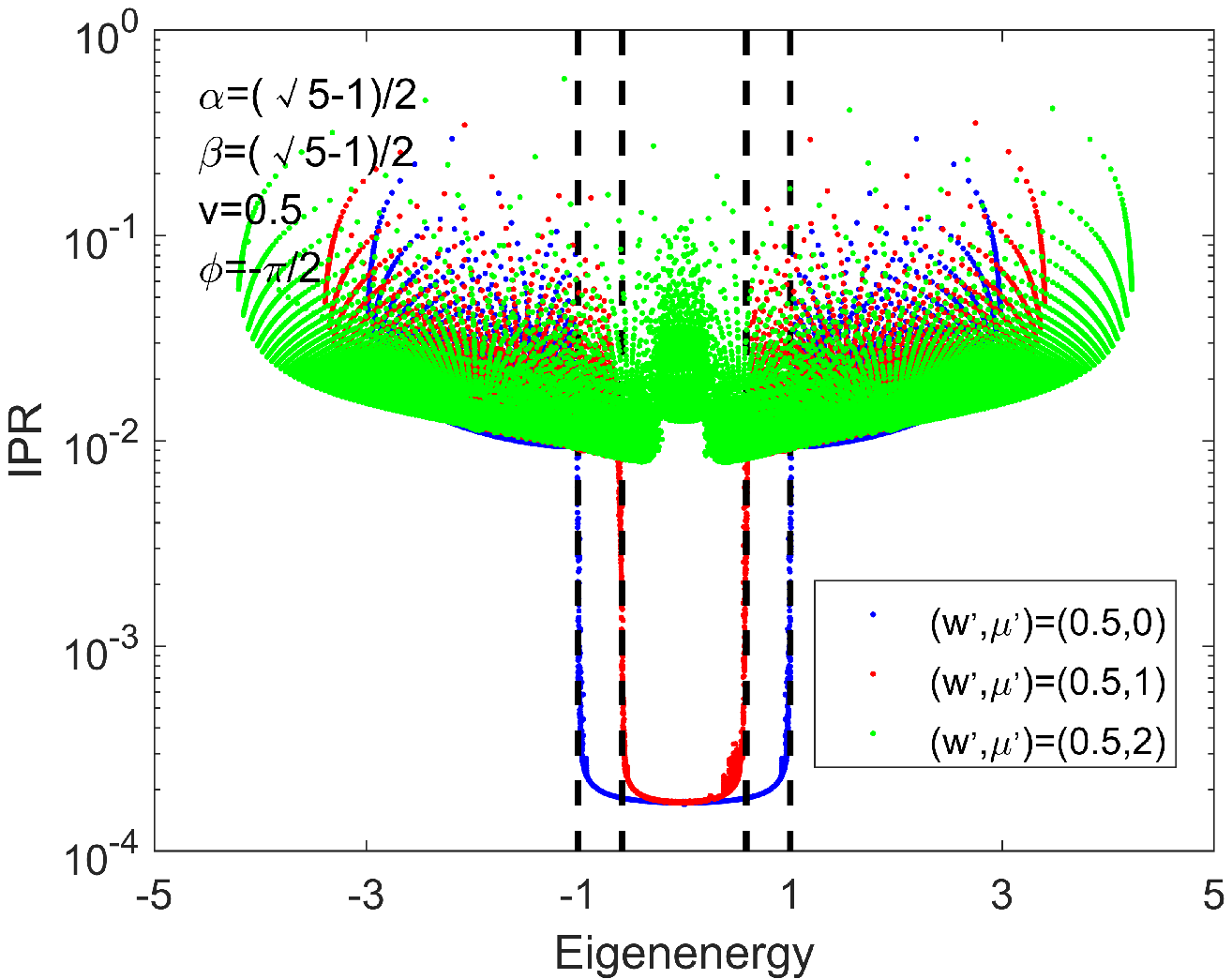}
		\end{minipage}
	}
	\centering
\caption{(a): Eigenenergy of AA model Eq.(\ref{AAH}) as a function of $\mu$ with $\alpha=\beta=(\sqrt{5}-1)/2$, $w=0.5$, $\phi=-\pi/2$and $M=200$.  (b): Eigenenergy of slowly varying AA model Eq.(\ref{AAH'}) as a function of $\mu'$ with $w'=w$ and $v=0.5$, other parameters are the same as (a). The colors in panels (a) and (b) represent the IPR of each eigenstates. The red and green curves represent the mobility edges. (c): The IPR of AA model as a function of eigenenergy for $(w,\mu)=(0.5,0),\,(0.5,1),\,(0.5,2)$, other parameters are the same as (a). (d): The IPR of slowly varying AA model with $w'=w$ and $v=0.5$, other parameters are the same as (c).}
\label{AAH_4}
\end{figure}

\textbf{AA model with different disorders in hopping and potential}

As the last example, we consider the case with different disorders in hopping and potential. This can be easily achieved by setting $\phi\neq0$. In this case, $w_j$ and $\mu_j$ are not the same disorder, but they are also not completely independent. To determine the mobility edges, we still consider a bunch of periodic models labeled by $a$ with $w_a=w\cos(2\pi\alpha a/M)$ and $\mu_a=\mu\cos(2\pi\beta a/M-\pi/2)=\mu\sin(2\pi\beta a/M)$. Here we set the phase difference to be $\phi=-\dfrac{\pi}{2}$. Since in the periodic model $a$,  $w_a$ and $\mu_a$ are only constants, we can simplify the Eq.(\ref{E1}) to the following
\be
E\in \bigcap_a\Big(R_{-a}-2t,\,R_{+a}+2t\Big)
\ee
here $R_a$ is given by
\be
R_{\pm a}=R\cos(\pm 2\pi\alpha\frac{a}{M}-\arctan(\dfrac{\mu}{2w})),\quad R=\sqrt{\mu^2+(2w)^2}
\ee
For a given $R$, the smallest interval is given by taking $R_{-a}=R$ for lower bound and $R_{+a}=-R$ for upper bound. Therefore, we find that the overlap of all intervals which is the region of extended states is given by
\be
E\in \Big(R-2t,\,-R+2t\Big)
\label{edge3}
\ee

To verify the above mobility edges, we plot the eigenenergy of AA mdoel with large quasi-period in hopping and potential in Figure \ref{AAH_4} (a). The parameters used in this calculation is listed in the figure caption. Again, the color of each points represents the IPR value of the eigenstates. As comparison, we also dusplay the eigenenergy of the slowly varying AA model eq.(\ref{AAH'}) with the same types of  disorders in hopping and potential in figure \ref{AAH_4} (b). It can be seen from the figure that the mobility edges of the two types of models are consistent with the calculated results of eq.(\ref{edge3}). In Figure \ref{AAH_4} (c), (d), we show the IPR values as a function of eigenenergy for a few selective points $\mu=0,\,1,\,2$. The panel (c) and (d) of Figure \ref{AAH_4} corresponds to the results of AA model with large quasi-period and slowly-varying AA model, respectively. It provides more quantitative evidences that supports the mobility edges of Eq.(\ref{edge3}). These figures also support the conclusion that the mobility edges of the two types of models are consistent.

\section{Qualitative explanations of the emergence of mobility edge with large quasi-period}
\label{sec-explain}

\begin{figure}[ht]
	\centering
	\subfigure[]{
		\begin{minipage}[t]{0.45\linewidth}
			\centering
			\includegraphics[width=\textwidth]{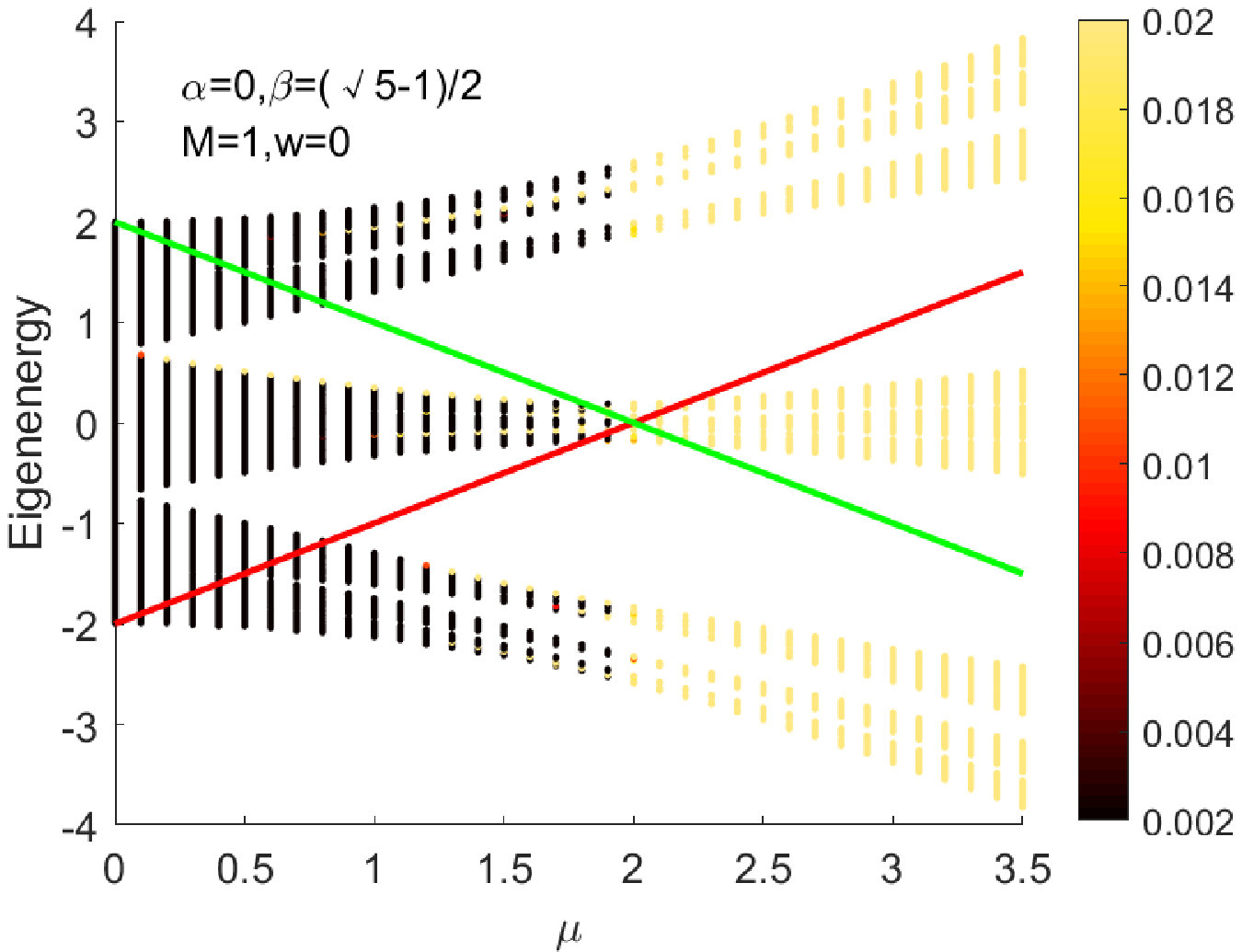}
		\end{minipage}
	}
	\subfigure[]{
		\begin{minipage}[t]{0.45\linewidth}
			\centering
			\includegraphics[width=\textwidth]{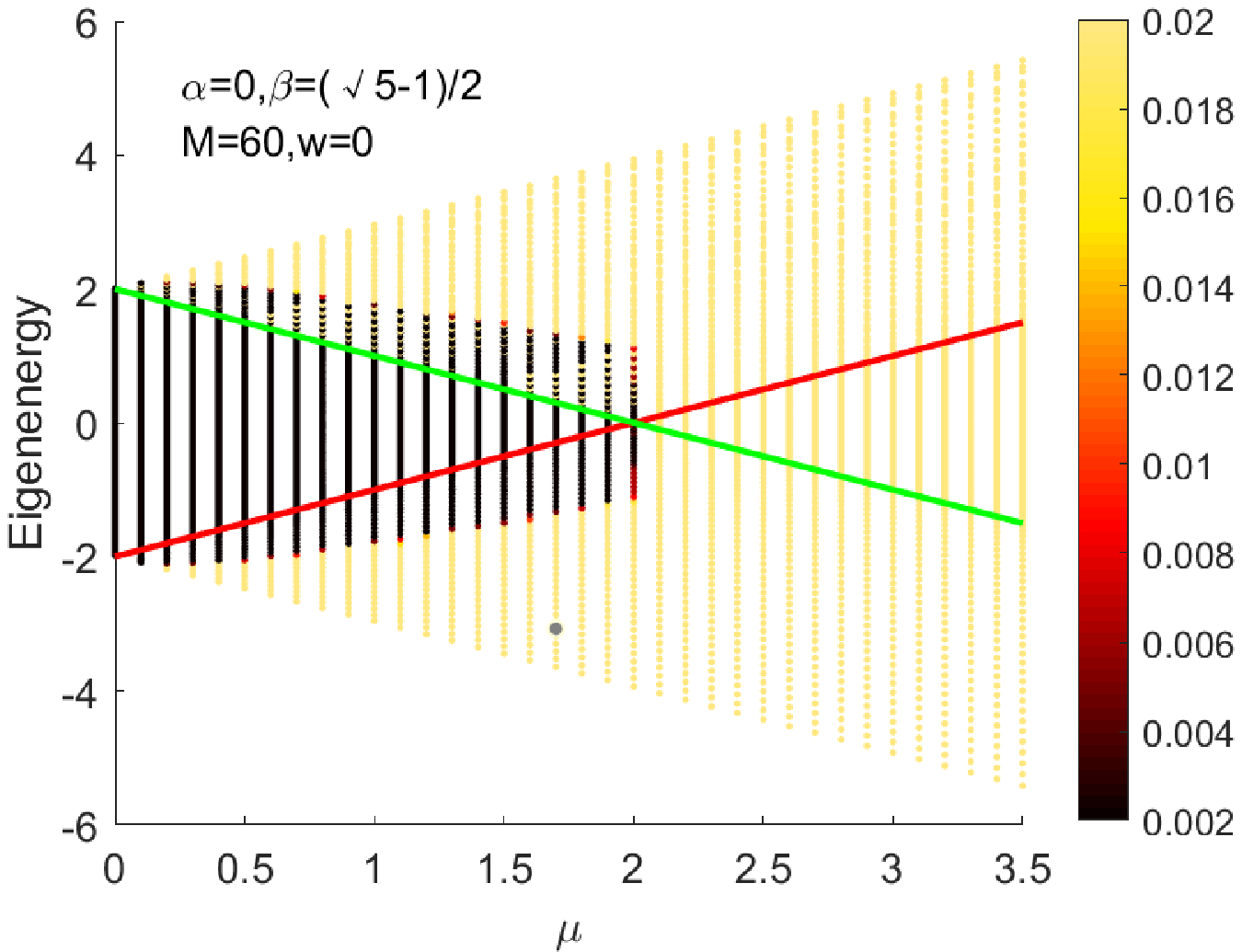}
		\end{minipage}
	}
	\quad
	\subfigure[]{
		\begin{minipage}[t]{0.45\linewidth}
			\centering
			\includegraphics[width=\textwidth]{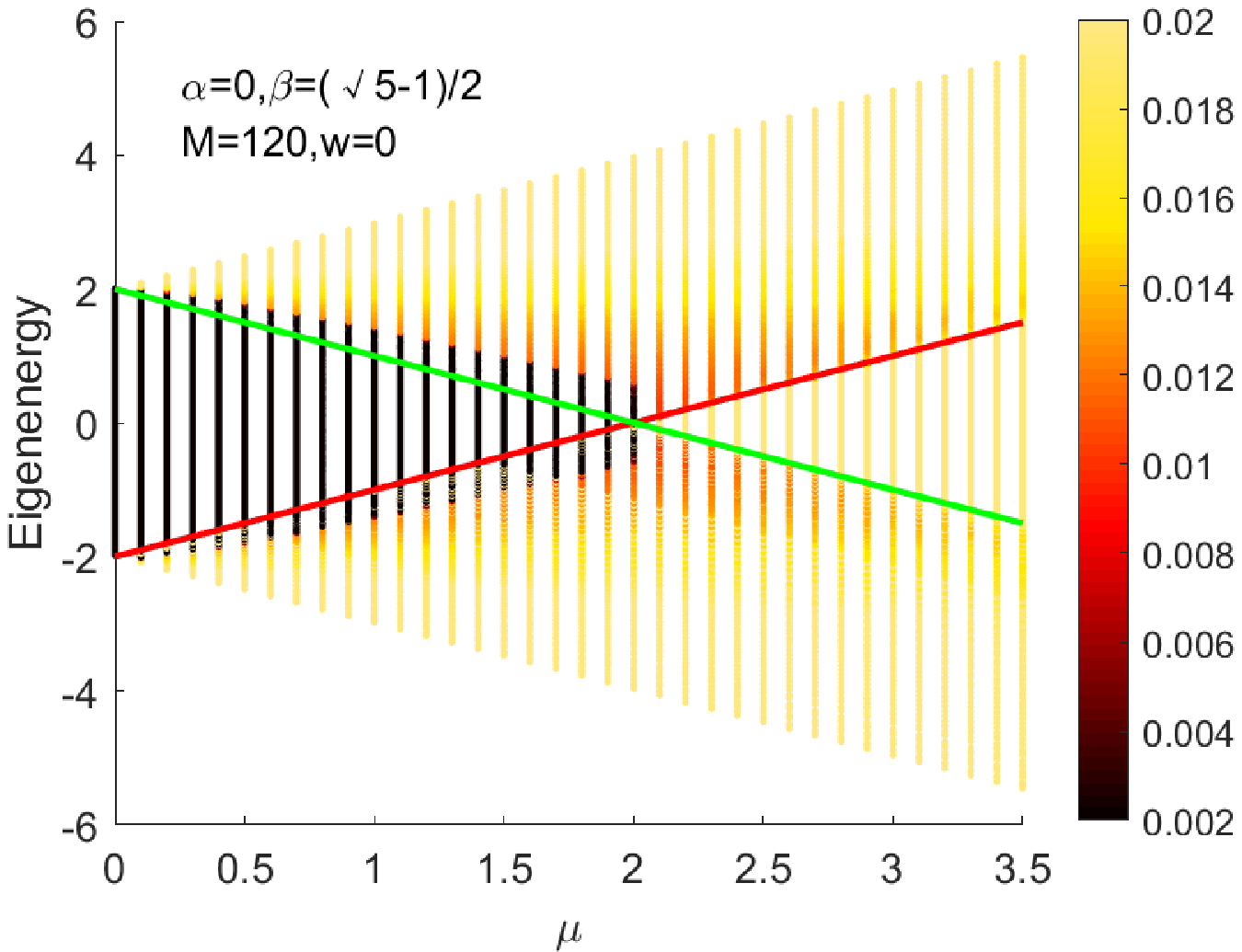}
		\end{minipage}
	}
	\subfigure[]{
		\begin{minipage}[t]{0.45\linewidth}
			\centering
			\includegraphics[width=\textwidth]{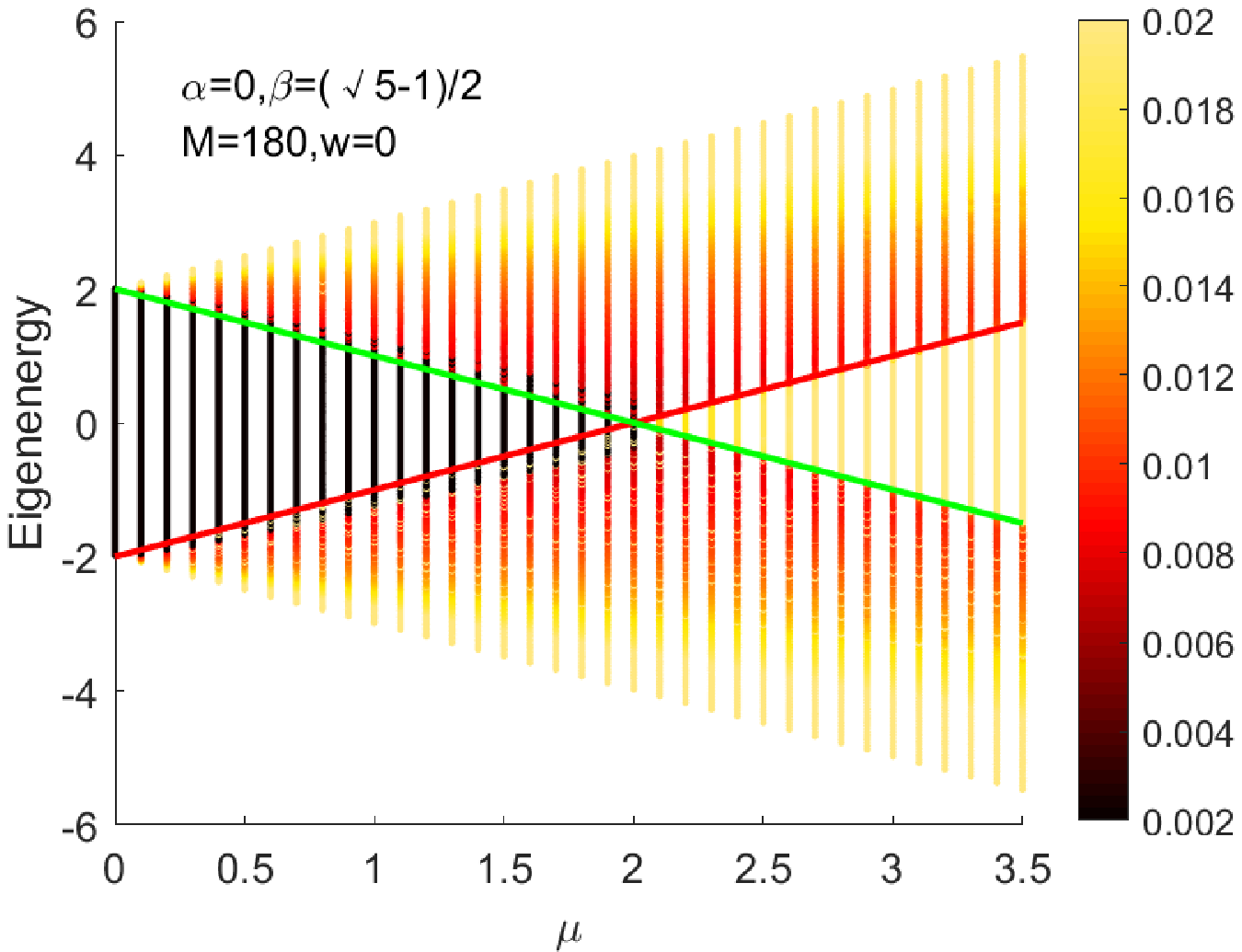}
		\end{minipage}
	}
	\centering
	\caption{Eigenenergy of AA model Eq.(\ref{AAH}) as a function of $\mu$ with $\alpha=0$, $\beta=(\sqrt{5}-1)/2$, $w=0$. The colors represent the IPR of each eigenstates. From (a) to (d), $M=1,\,60,\,120,\,180$ respectively.}
	\label{AAH_5}
\end{figure}

In the original AA model, there is a duality point. The eigenstates are all localized in one side of this point and are all extended in the other side. Because of this, there is no mobility edge exist in the original AA model. Our modification of AA model is to introduce a large integer to the make a large quasi-period $\alpha\to\dfrac{\alpha}{M}$. The large value of $M$ is crucial here to make the mobility edge to appear. Here we want to show the effects of $M$ on the model of Eq.(\ref{AAH}).

In Figure \ref{AAH_5}, we plot the eigenenergy of AA model of Eq.(\ref{AAH}) with different $M$ values. The parameters used in this calculation is listed in the figure caption. The color of each points represents the IPR value of the eigenstates. In panel (a), we have $M=1$ which is exactly the original AA model. The duality point is located at $\mu=2t$. One can clearly see that the extended and localized states are separated by this duality point. Therefore, for a fixed $\mu$, there is no mobility edges. From panel (b) to (c), we gradually increase from $M=60$ to $M=120$. Then for a fixed $\mu$, there gradually appears more and more localized states. But the location of mobility edge is still not very clear.  When we reach $M=180$, the mobility edge becomes quite sharp and its location is the same as the mobility edge of the slowly varying AA model. These figures clearly shows the relationship between the size of $M$ and the existence of mobility edge.

To qualitatively understand why the mobility edges can exist in the large quasi-periodic models, we can still think about the slowly varying models first. In the slowly varying models, the disordered term $\mu_i$ become constant as the site number $i$ become very large. Clearly, the constant $\mu_i$ supports extended states, while disordered $\mu_i$ support localized states. The competing of these two effects leads to the existence of mobility edges. Similar arguments also works for the large quasi-periodic models. Since for very large $M$, there are also many $\mu_i$ has almost the same value. Thus, they also look like constant potential that supports extended states. This will also lead to the competing between extended and localized states, which in turn give rise to the appearance of mobility edge. This also requires large enough $M$ to ensure there are sufficient number of $\mu_i$ taking similar values.

\begin{acknowledgements}
This work was supported by the Natural Science Foundation  of  China  under  Grant  No.   11874272  and  Science Specialty  Program  of  Sichuan  University  under  Grant No. 2020SCUNL210.
\end{acknowledgements}

\end{document}